\documentclass[numberedappendix, iop, apj]{emulateapj}
\usepackage{apjfonts}
\usepackage{natbib,natbibspacing}
\usepackage{subfigure}

\usepackage{etex}
\reserveinserts{18}
\usepackage{morefloats}

\def\apj{ApJ}

\def\mnras{MNRAS}

\def\araa{ARAA}
\def\aj{AJ}
\def\aap{A $\&$ A}

\def\apjs{ApJS}
\def\pasp{PASP}

\def\lya{Ly-$\alpha$}

\def\nh{$N_{\rm H I}$}
\def\Z{$\langle {\rm Z}\rangle$}

\shorttitle{Metallicity of DLAs at $z\sim5$}
\shortauthors{Rafelski, Neeleman, Fumagalli, Wolfe, \& Prochaska}

\begin{document}

\title{The Rapid Decline in Metallicity of Damped {\lya} Systems at {$\MakeLowercase{z}\sim5$}} 

\author{Marc Rafelski\altaffilmark{1}, Marcel Neeleman\altaffilmark{2}, Michele Fumagalli\altaffilmark{3,4,5}, Arthur M. Wolfe\altaffilmark{2}, J. Xavier Prochaska\altaffilmark{6}, }

\altaffiltext{1}{Infrared Processing and Analysis Center, Caltech, Pasadena, CA 91125, USA}
\altaffiltext{2}{Department of Physics and Center for Astrophysics and Space Sciences, UCSD, La Jolla, CA 92093, USA}
\altaffiltext{3}{Carnegie Observatories, 813 Santa Barbara Street, Pasadena, CA 91101, USA.}
\altaffiltext{4}{Department of Astrophysics, Princeton University, Princeton, NJ 08544-1001, USA.}
\altaffiltext{5}{Hubble Fellow.}
\altaffiltext{6}{Department of Astronomy \& Astrophysics,  UCO/Lick Observatory, 1156 High Street, University of California, Santa Cruz, CA 95064}

\begin{abstract} 
We present evidence that the cosmological mean metallicity of neutral atomic hydrogen gas shows a sudden decrease at $z>4.7$ down to $\langle {\rm Z}\rangle=-2.03^{+0.09}_{-0.11}$,
which is  $6\sigma$ deviant from that predicted by a linear fit to the data at lower redshifts. This measurement is made possible by the chemical abundance measurements of
8 new damped Ly-$\alpha$ (DLA) systems at $z>4.7$ observed with the Echellette Spectrograph and Imager on the Keck II telescope, 
doubling the number of measurements at $z>4.7$ to 16. 
Possible explanations for this sudden decrease in metallicity include a change in the physical processes that enrich the neutral gas within disks, or an increase of the covering factor of neutral gas outside disks due to a lower ultra-violet radiation field and higher density at high redshift.
The later possibility would result in a new population of presumably lower metallicity DLAs, with an increased contribution to the DLA population at higher redshifts resulting in a reduced mean metallicity.
Furthermore, we provide evidence of a possible decrease at $z>4.7$ in the comoving metal mass density of DLAs, $\rho_{\rm metals}(z)_{\rm DLA}$, which is flat out to $z\sim4.3$. 
Such a decrease is expected, as otherwise most of the metals from star-forming galaxies would reside in DLAs by $z\sim6$.
While the metallicity is decreasing at high redshift, the contribution of DLAs to the total metal budget of the universe increases with redshift, with DLAs at $z\sim4.3$ accounting for $\sim20$\% as many metals as produced by Lyman break galaxies.
\end{abstract}


\clearpage

\section{Introduction}

Damped {\lya} systems (DLAs) are atomic hydrogen gas clouds measured in absorption to background quasars with a minimum column 
density of $2\times10^{20}$cm$^{-2}$ that dominate the neutral-gas content of the Universe at high redshift \citep{Wolfe:2005}. 
Simulations and semi-analytic models predict that the majority of gas that gives rise to DLAs is associated with galaxies \citep[e.g.][]{Nagamine:2004b, Razoumov:2008, Fumagalli:2011,Cen:2012, Berry:2013}, 
and this picture is increasingly supported by observations \citep[e.g.][]{Rafelski:2011, Fynbo:2013, Peroux:2013}.

DLA metal abundances are the most robust measures of metallicity yet established at high redshift.
While it is difficult to measure the chemical properties of faint star-forming galaxies measured in emission, they are more easily determined from the gas detected in absorption to bright background quasars \citep[e.g.][]{Prochaska:2003a}. 
Moreover, because DLAs are mainly neutral, ionization corrections are not needed to determine elemental abundances, as is the case 
for star-forming galaxies \citep[e.g.][]{Erb:2006}, the {\lya} forest \citep[e.g.][]{Aguirre:2008}, and Lyman Limit Systems \citep[e.g.][]{Lehner:2013}.

We therefore use DLAs to measure the cosmic metallicity of neutral hydrogen, {\Z},  at $z\sim5$, where 
$\langle{\rm Z}\rangle=\log(\Omega_M/\Omega_{HI})-\log(\Omega_M/\Omega_{HI})_\odot$,
and $\Omega_M$ and $\Omega_{HI}$ are the comoving densities of metals and of atomic hydrogen \citep{Lanzetta:1995}.
Previously, {\Z} was determined out to $z\sim4.7$, with a $\sim7\sigma$ significant  decline in {\Z} with increasing redshift, given by 
$\langle{\rm Z}\rangle=(-0.22{\pm}0.03)z-(0.65{\pm}0.09)$  \citep[][hereafter R12]{Rafelski:2012}, confirming previous $\sim3\sigma$
detections \citep{Prochaska:2003a, Kulkarni:2005,Kulkarni:2010}. 
One of the most intriguing results from R12 is that they find
a rapid decrease in metallicity at $z>4.7$, although with only 7 measurements. 
In this letter we double the sample of $z>4.7$ metallicities to investigate a possible break in the evolution of {\Z}. 

Having established the metallicity of DLAs over 12 Gyr of cosmic evolution, we investigate the contribution of DLAs to the total metal content of the Universe 
by examining the comoving mass density of metals, 
$\rho_{\rm metals}$, for DLAs compared to that produced in star-forming regions of galaxies.
While DLAs contribute only a small fraction of the total metal budget of the universe at $z\sim2$ \citep{Pettini:2004, Pettini:2006, Bouche:2007}, 
their contribution may increase at higher redshifts.

Throughout this paper we adopt  a  cosmology with (${\Omega_{\rm M}},{\Omega_{\Lambda}},h$)=(0.3,0.7.0.7). 

\section{Observations and Measurements}
\label{obs}

We use the Echellette Spectrograph and Imager \citep[ESI;][]{Sheinis:2002} on the Keck II telescope,
with a FWHM of $\sim44$ km~s$^{-1}$ (with 0.75$\tt''$ slit, $R\approx7,000$), to observe 
28 $z>4.7$ quasars containing 42 candidate DLAs.
The DLAs are selected from the following low resolution spectroscopic surveys of $z\sim5$ quasars: 11 candidates from 
the Gemini/GMOS survey \citep{Worseck:2013}, 30 from the SDSS-DR9 survey  \citep{Ahn:2012,Noterdaeme:2012}, 
and 1 from the SDSS-DR10 survey \citep{Ahn:2013} as found by our team.
The new observations comprise a total of 5 nights on ESI, obtained in March 2012, January 2013, May 2013, and August 2013.

\begin{figure}
\center{
\includegraphics[angle=0,scale=0.5, viewport=30 10 520 620,clip]{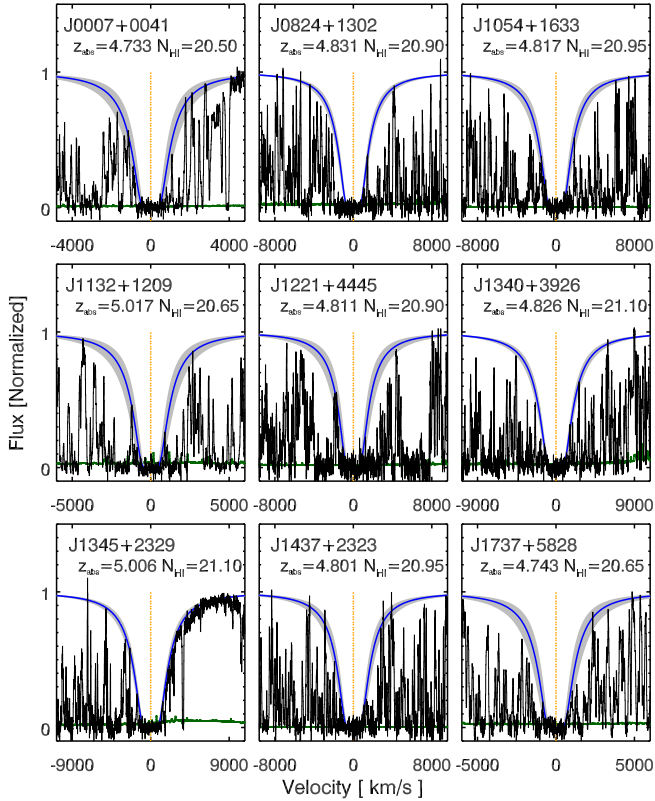}
}
\caption{Voigt profile fits for the 9 confirmed DLAs at $z>4.7$, with the x-range based on their {\nh}. Blue curves are best-fit 
profiles and gray shade includes 95\% confidence limits surrounding best fits.
The orange dotted line marks the velocity centroid of the DLA determined from the low-ion metal transitions.
The green line represents the uncertainty.
}
\label{fig:nhi}
\end{figure}

We follow the same observing, data reduction, and analysis methodology as described in R12.  
We fit {\nh} Voigt profiles to the {\lya} lines of 42 candidate $z>4.7$ DLAs and confirm 9 as bona-fide DLAs. We also confirm 
8 $z<4.7$ DLAs in the spectra. 
Figure \ref{fig:nhi} shows the Voigt profile fits of the confirmed $z>4.7$ DLAs. 

We test the quality of the {\nh} fits by simulation, where we plant artificial DLAs into actual sight-lines and into realistic mock spectra. 
The simulations yield a mean offset of 0.02 and 0.06 dex, with standard deviations of 0.07 and 0.12 dex for the mock spectra and 
actual sight-lines respectively. These values can be compared to the conservative errors of 0.1-0.2 dex quoted in 
Table \ref{metal:tab}, which shows that our {\nh} fits are robust.

The misidentification rate of DLAs at $z>4.7$ is significantly higher than at lower redshift.
Of the 30 $z>4.7$ candidate DLAs from the SDSS-DR9 sample \citep{Noterdaeme:2012},
we confirm only 1 DLA, for a misidentification rate of $\sim97$\%. For the 11 $z>4.7$ candidates from the GMOS survey,
which has significantly higher S/N and used manual candidate selection, we confirm 7 DLAs for a misidentification rate of $\sim36$\%. 

The higher misidentification rate is due to a combination of the increasing density of the {\lya} forest at $z>4.7$, 
and fainter quasars resulting in lower S/N in the SDSS spectra. 
While the spectral resolution of the SDSS surveys of FWHM $\sim2$\AA ~is sufficient to find DLAs with a high degree of confidence at $z<4$, 
an echellette on a 10m class telescope is optimal to resolve out the {\lya} forest, and reliably measure {\nh} for DLAs at $z>4.7$.

We measure metal abundances of DLAs with the same methodology as R12, using various elements to determine the 
metal abundances based on availability of clean unblended and unsaturated lines in the spectrum. 
At the resolution of ESI, narrow lines are not typically saturated at a normalized intensity of $F_{min}/F_q \gtrsim 0.5$, 
where $F_{min}$ is the minimum flux and $F_q$ is the quasar flux \citep[R12,][]{HerbertFort:2006,Penprase:2010,Jorgenson:2013}.
In addition, we consider ratios of multiple lines, when available, to check for possible saturation and blending.
Only unsaturated lines are used, yielding metallicities without the need for any saturation corrections. 

\begin{figure}
\center{
\includegraphics[angle=0,scale=0.56, viewport=70 315 520 610,clip]{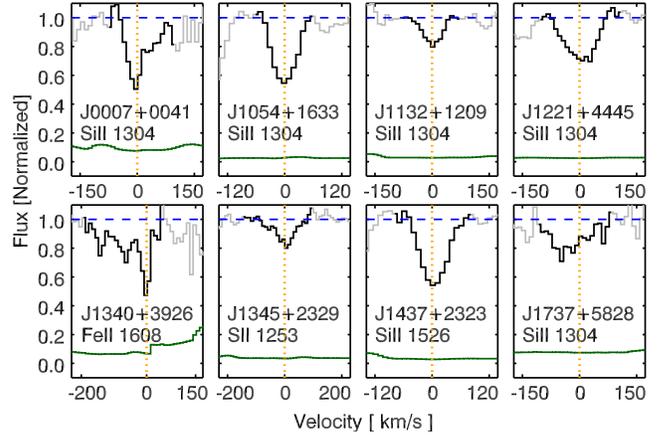}
}
\caption{Primary metal transitions for 8 new  $z>4.7$ DLAs with metallicity measurements, with the x-range based on the transition velocity width.
The black regions mark the used velocity width of the transitions for determining the metallicity, 
the orange dotted line marks the velocity centroid of the DLA, the blue dashed line marks the continuum level, and the green line represents the uncertainty.
Three DLAs have their metallicity determined from more than one metal transition, but only one example transition is shown here. 
}
\label{fig:trans}
\end{figure}

We present new metallicity measurements for 16 DLAs in Table \ref{metal:tab}, with 8 of them at $z>4.7$.
We note that one DLA does not have measurable metallicities because all clean non-blended detected lines are saturated, 
and list it as a limit in the table.
Figure \ref{fig:trans} shows the metal transitions used to determine metal abundances for 8 of the 9 new DLAs at $z>4.7$.
The velocity range used for each DLA is determined from other saturated transitions, while avoiding artifacts or unrelated absorption,
shown in black in Figure \ref{fig:trans}.

In addition to the new metallicities, we use the 241 DLA metallicities compiled by R12. 
To this we add one new DLA from the literature, J1208+0010,  observed with X-Shooter \citep{DOdorico:2006} on the Very Large Telescope
with a resolution of $\sim30$ km s$^{-1}$  \citep{Becker:2012}. 
We also include an upper limit for J0824+1302 from preliminary data from FIRE \citep{Simcoe:2008} observations on the Magellan Baade telescope, which will be presented in upcoming work. 
We do not include the new sample by \citet{Jorgenson:2013}, as it
does not include any $z>4.5$ DLAs. 
In total, this results in a sample of 258 DLA metallicities. 

\begin{deluxetable*}{rrrrcrcrcrc}
\tabletypesize{\scriptsize}
\tablecaption{New DLA metallicities
\label{metal:tab}}
\tablewidth{0pt}
\tablehead{
\colhead{QSO} &
\colhead{$z_{em}$} &  
\colhead{$z_{abs}$} &  
\colhead{log$N_{\rm HI}$} & 
\colhead{$f_{\rm \alpha}$\tablenotemark{a}} &
\colhead{[$\alpha$/H]} &
\colhead{$f_{\rm Fe}$\tablenotemark{b}} &
\colhead{[Fe/H]} & 
\colhead{$f_{\rm mtl}$\tablenotemark{c}} &
\colhead{[M/H]}
}

\startdata
J0007+0041 &4.828 &4.7333 &20.50$ \pm $0.20 &1 &$-2.04\pm $0.20 &0 & \nodata &1 &$-2.04\pm $0.20 \\
J0824+1302 &5.188 &4.8308 &20.90$ \pm $0.10 &2 &$>-2.24\pm $0.10 &3 &$<-2.12\pm $0.10\tablenotemark{d} &0 & \nodata \\
J0824+1302 &5.188 &4.4720 &20.30$ \pm $0.10 &1 &$-1.97\pm $0.12 &1 &$-2.15\pm $0.13 &1 &$-1.97\pm $0.12 \\
J1054+1633 &5.187 &4.8166 &20.95$ \pm $0.15 &1 &$-2.47\pm $0.15 &0 & \nodata &1 &$-2.47\pm $0.15 \\
J1054+1633 &5.187 &4.1346 &21.00$ \pm $0.10 &1 &$-0.70\pm $0.11 &0 & \nodata &1 &$-0.70\pm $0.11 \\
J1054+1633 &5.187 &3.8420 &20.60$ \pm $0.20 &0 & \nodata &1 &$-2.47\pm $0.21 &2 &$-2.17\pm $0.18 \\
J1132+1209 &5.167 &5.0165 &20.65$ \pm $0.20 &1 &$-2.66\pm $0.20 &3 &$<-2.55\pm $0.20 &1 &$-2.66\pm $0.20 \\
J1132+1209 &5.167 &4.3802 &21.20$ \pm $0.20 &0 & \nodata &1 &$-2.87\pm $0.21 &2 &$-2.57\pm $0.17 \\
J1204$-$0021 &5.090 &3.6444 &20.70$ \pm $0.10 &0 & \nodata &1 &$-2.30\pm $0.11 &2 &$-2.00\pm $0.17 \\
J1208+0010 &5.270 &5.0817 &20.30$ \pm $0.15 &1 &$-2.06\pm $0.15\tablenotemark{e} &1 &$-2.48\pm $0.17\tablenotemark{e} &1 &$-2.06\pm $0.15\tablenotemark{e} \\
J1221+4445 &5.206 &4.8110 &20.90$ \pm $0.20 &1 &$-2.46\pm $0.20 &1 &$-2.01\pm $0.21 &1 &$-2.46\pm $0.20 \\
J1245+3822 &4.940 &4.4470 &20.85$ \pm $0.10 &1 &$-2.34\pm $0.11 &3 &$<-2.37\pm $0.10 &1 &$-2.34\pm $0.11 \\
J1340+3926 &5.026 &4.8264 &21.10$ \pm $0.10 &2 &$>-2.23\pm $0.10 &1 &$-2.22\pm $0.11 &2 &$-1.92\pm $0.17 \\
J1345+2329 &5.119 &5.0060 &21.10$ \pm $0.10 &4 &$-1.59\pm $0.11 &0 & \nodata &1 &$-1.59\pm $0.11 \\
J1418+3142 &4.850 &3.9625 &20.90$ \pm $0.15 &1 &$-0.64\pm $0.15 &4 &$-0.72\pm $0.15 &1 &$-0.64\pm $0.15 \\
J1437+2323 &5.320 &4.8007 &20.95$ \pm $0.15 &1 &$-2.64\pm $0.15 &0 &\nodata &1 &$-2.64\pm $0.15 \\
J1626+2858 &5.022 &4.6078 &20.30$ \pm $0.15 &1 &$-2.38\pm $0.18 &0 & \nodata &1 &$-2.38\pm $0.18 \\
J1737+5828 &4.941 &4.7435 &20.65$ \pm $0.20 &1 &$-2.23\pm $0.21 &0 & \nodata &1 &$-2.23\pm $0.21 
\enddata

\tablecomments{None of the reported limits take into account the uncertainty in $N_{\rm H I}$. The abundance uncertainties are dominated by the uncertainty in $N_{\rm H I}$. }
\tablenotetext{a}{0 = No measurement; 1= Si measurement; 2 = Si lower limit; 3 = Si upper limit; 4 = S measurement, .}
\tablenotetext{b}{0 = No measurement; 1= Fe measurement; 2 = Fe lower limit; 3 = Fe upper limit; 
4 = [Ni/H]$-$0.1 dex}
\tablenotetext{c}{0 = No measurement; 1 = [$\alpha$/H]; 2=[Fe/H]. In the latter case, we use [M/H] = [Fe/H]+0.3 dex.}
\tablenotetext{d}{Preliminary limit from FIRE observations not presented in this letter. }
\tablenotetext{e}{Measurement from \citet{Becker:2012}. }

\end{deluxetable*}

\begin{figure*}
\center{
\includegraphics[scale=0.5, viewport=15 5 495 360,clip]{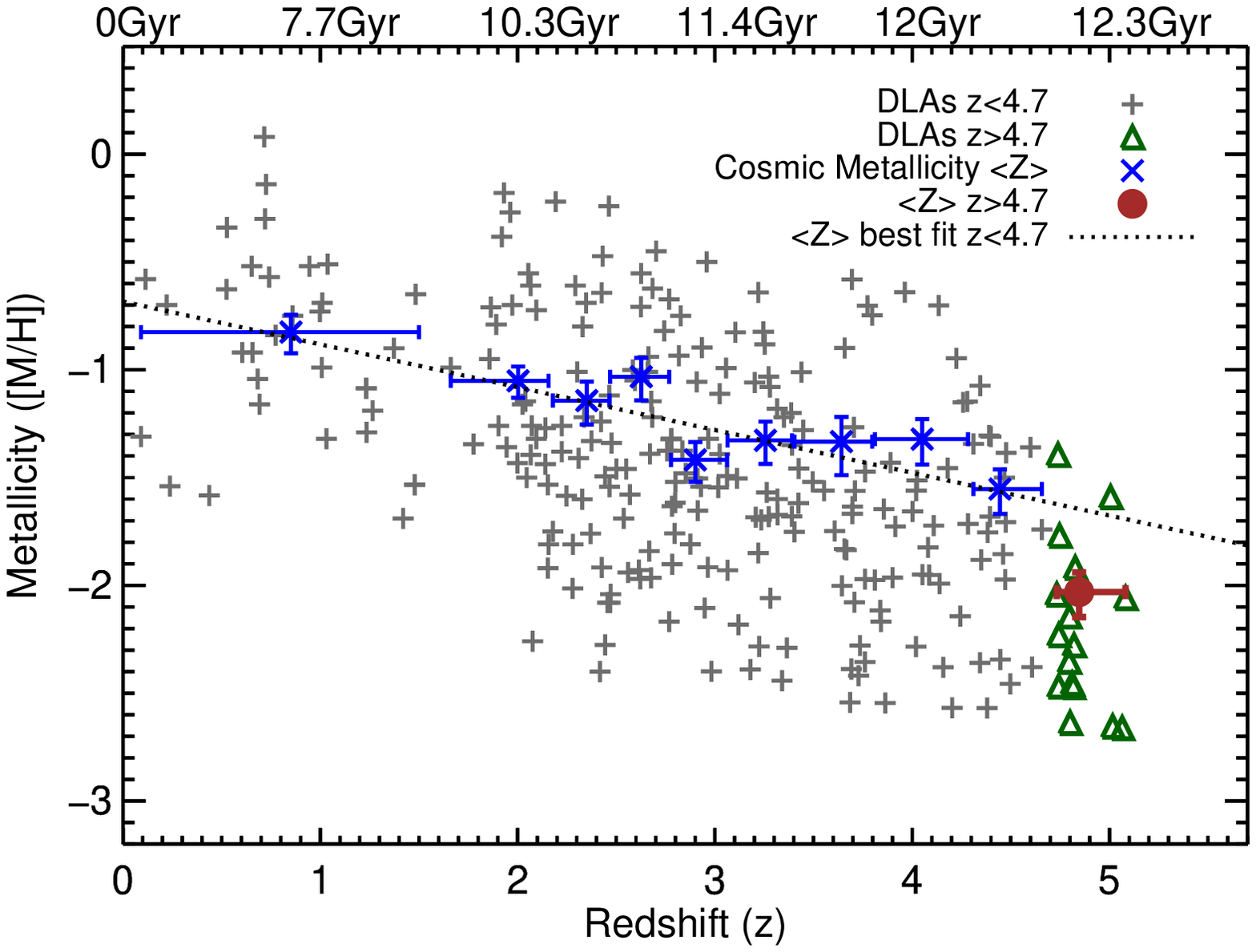}
\includegraphics[scale=0.5, viewport=15 5 495 360,clip]{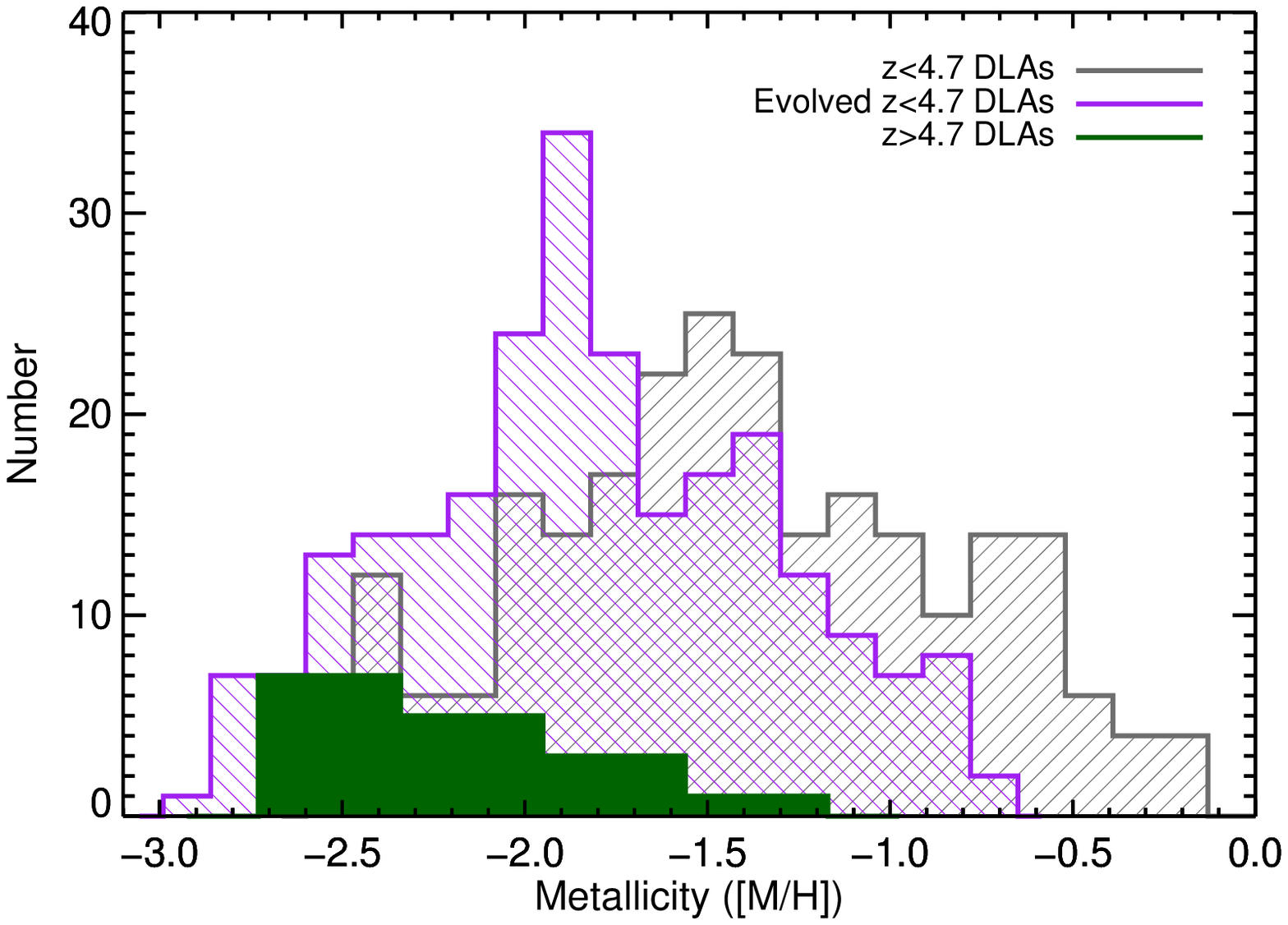}
}
\caption{
{\bf Left:} DLA metallicity versus redshift, showing a sharp decrease in metallicity at $z>4.7$.
The gray plus signs are metallicities of DLAs at $z<4.7$, and the green triangles are DLAs at $z>4.7$.
The blue crosses show the cosmic metallicity, {\Z}, with the vertical error bars 
representing 1$\sigma$ confidence levels from our bootstrap analysis. 
The black dotted line is a linear fit to the {\Z} data points in redshift space for DLAs $z<4.7$.
The brown circle is {\Z} deduced from DLAs at $z>4.7$, and is significantly below the linear fit.
{\bf Right:} Histogram of DLA metallicities, showing a very different metallicity distribution of $z>4.7$ DLAs.
The gray region represents the $z<4.7$ DLAs, the purple region corresponds to $z<4.7$ DLAs evolved to $z=4.85$, and the filled green region 
is for $z>4.7$ DLAs.}
\label{fig:metal}
\end{figure*}

\section{Metallicity Evolution}

The evolution of heavy elements in neutral gas can be described by the cosmological mean metallicity {\Z}.
While we cannot measure the cosmological values of $\Omega_M$ and $\Omega_{HI}$ directly, we can 
compute {\Z} from a column density weighted mean of individual systems, as described by equation 1 in R12.
In this way, {\Z} was found to be decreasing with increasing redshift out to $z\sim4.7$ (R12).  
We plot the metallicity as a function of redshift for the 258 DLA abundances described in \S\ref{obs} in the left panel of Figure \ref{fig:metal}, 
with the corresponding {\Z} measurements plotted as blue points. 
The {\Z} statistic is dominated by systems with the highest {\nh}, and thus the uncertainties are dominated by sample variance rather than statistical error, and are calculated via a bootstrap method as described in R12.
The large dispersion in the left panel of Figure \ref{fig:metal} is not observational error, but is due to intrinsic scatter caused by the different DLA host galaxy masses \citep{Neeleman:2013}. 
Its magnitude requires a substantial sample of DLAs per bin to provide an accurate estimate of {\Z}, and $\sim25$ DLAs per bin is typically sufficient for an accurate measurement (R12).

In addition to the metallicity evolution, R12 found an apparent drop in metallicity at $z>4.7$, although based on only 7 measurements. 
Here we increase the number of metallicity measurements at $z>4.7$ by 8, for a total of 16 (including one new measurement from \citet{Becker:2012}).
In order to determine if there is a deviation in the mean metallicity of DLAs at $z>4.7$, we fit a line to the {\Z} values up to $z=4.7$.
The $z>4.7$ DLAs are not included, as this is the interval we want to test variation for. 
We find that $\langle{\rm Z}\rangle=(-0.20{\pm}0.03)z-(0.68{\pm}0.09)$, consistent with R12. 

Of the 16 DLAs at $z>4.7$, 14 have metallicities below the extrapolated linear fit ({\Z}$=-1.65$ at $z=4.85$).
A column density weighted mean of these 16 DLAs yields $\langle {\rm Z}\rangle=-2.03~^{+0.09}_{-0.11}$ at $z\sim4.85$, which is a $6\sigma$ deviation from the extrapolated linear fit in linear space.
We test the possibility of small number statistics yielding the above result via Monte Carlo simulations. 
The metallicities of the $z<4.7$ sample are evolved out to the mean redshift of the $z>4.7$ sample ($z=4.85$) by adding the difference in metallicity in the above fit
at $z=4.85$ and at each DLA redshift. If the {\nh} values are kept the same, then the probability of drawing a sample of 16 DLAs from the evolved $z<4.7$ sample with the same or lower  {\Z}
is 1.0\%. If {\nh} is randomly sampled, it is 1.1\%. 

We also compare the distribution of metallicities
using the Kolmogorov-Smirnov (K-S) test to find the probability that the $z>4.7$ sample 
distribution is drawn from the same parent population as the evolved $z<4.7$ sample. The probability is 0.6\%, and therefore the hypothesis that the two samples are drawn from the same population
can be rejected at almost $3\sigma$ confidence. The metallicity distributions are compared in the right panel of Figure \ref{fig:metal}. We repeat the same test for the solar normalized metal column density 
distributions rather than metallicity, and find that the probability of the two being drawn from the same population is also rejected at $3\sigma$ confidence.  This shows that lower metal column densities is driving the lower metallicities at $z>4.7$, rather than {\nh}.

One possible concern is that since {\Z} is a column density weighted mean, a few high {\nh} measurements can affect {\Z}. 
The {\nh} distribution of $z>4.7$ DLAs matches that from the SDSS survey \citep{Prochaska:2009}, with a K-S probability of 74\%. 
We check for the possibility that we have a non-standard distribution of {\nh} as a function of the metallicity by Monte Carlo simulations.
{\Z} is calculated from the 16 $z>4.7$ metallicities, and randomly sampling {\nh} from the SDSS {\nh} distribution. The resultant
distribution is consistent with our measured {\Z}, with $\bar{\langle {\rm Z}\rangle}=-2.04\pm0.11$. Alternatively, a simple average of the $z>4.7$ metallicities 
yields a metallicity of [M/H]$=-2.02$. This shows that the distribution of {\nh} is not driving our results, while the metallicity distribution is. 

We also check if the one DLA for which we do not have a robust metallicity biases our sample low. We find that
if we include the the upper limit in Table \ref{metal:tab} as an actual measurement, we obtain a slightly smaller {\Z}, and therefore excluding this DLA from the sample does not significantly affect our result. 

There is nothing special about $z=4.7$, but because $z>4.7$ DLAs were specifically targeted after indications of a possible break at $z>4.7$ in R12, 
observations are lacking between $z=4.5-4.7$.
If instead the 18 DLAs at $z>4.5$ are selected, then {\Z}$= -1.98~^{+0.08}_{-0.10}$, which is very similar to the $z>4.7$ result.
In order to move {\Z} within $1\sigma$ of the extrapolated fit, the cut would need to be shifted to $z>4.34$.
However, as this is a transition occurring at high redshift, it makes sense to use a cut at the highest redshift that maintains sufficient numbers.

These tests show that a model of linear evolution in {\Z}, which describes the data well at $z<4.5$, is ruled out at higher redshifts by the current dataset.
This indicates that a rapid change in the average metallicity of neutral HI gas occurs beyond $z\sim4.7$.

\section{Mass Density of Metals}
\label{metaldensity}

Having measured the metallicity of DLAs out to $z\sim5$, we can consider the contribution of DLAs to the metal budget of the universe as a function of redshift. 
We do so by comparing the comoving metal mass density, $\rho_{\rm metals}(z)$, in DLAs to the metals produced in star-forming regions of galaxies. 
 We calculate  $\rho_{\rm metals}(z)$ for DLAs as  
 \begin{equation}
\rho_{\rm metals}(z)_{\rm DLA} =  10^{\langle {\rm Z}\rangle} \times \rho_{\rm HI} \times (Z/X)_\odot \;,
\label{eq:rhometaldla}
\end{equation}

 \noindent~where $\rho_{\rm HI}$ is the comoving \ion{H}{1} mass density \citep{Prochaska:2009, Noterdaeme:2009, Noterdaeme:2012},
 and $(Z/X)_\odot$ is the solar normalization, $(Z/X)_\odot=0.0181$ \citep{Grevesse:2010}. Here, we use $\rho_{\rm HI}$ from \citet{Prochaska:2009} as it extends to higher redshifts, and recalculate {\Z}  using the same redshift bins as in \citet{Prochaska:2009}. We also calculate $\rho_{\rm metals}(z)_{\rm DLA}$ for our $z>4.7$ sample, for which we need to linearly extrapolate the $\rho_{\rm HI}$ measurements to the  mean redshift to obtain a measurement of $\rho_{\rm metals}(z)_{\rm DLA}$ at $z=4.85$. The reader should note that this extrapolation bears significant uncertainty, as it assumes that $\rho_{\rm HI}$  continues to increase with redshift.

While the total metal mass density of the universe, $\rho_{\rm metals}(z)_{\rm U}$, is difficult to quantify \citep{Peeples:2013}, the production of metals at high redshift is dominated by Lyman break galaxies (LBGs) \citep{Bouche:2005, Bouche:2006, Bouche:2007, Pettini:2006}. LBGs provide a lower limit to the total metals produced, accounting for about half of the metals \citep{Pettini:2006, Bouche:2007}, and thus are a good proxy for the production of metals in the Universe.
We calculate $\rho_{\rm metals}(z)$ produced by LBGs per redshift interval by integrating the star formation rate density, $\dot{\rho_{*}}$, of LBGs \citep{Reddy:2009, Bouwens:2012,Oesch:2013c} to obtain the comoving mass density of stars in LBGs, $\dot{\rho_*}$, and multiplying by the estimated conversion factor for the metal production rate ${\rho}_{\rm metals} = (1/64)\dot{\rho_*}$ \citep{Conti:2003, Pettini:2006}. Specifically, 

\begin{equation}
\rho_{\rm metals}(z)_{\rm LBG}=\frac{1}{64} \times\int^{z^\prime_2}_{z^\prime_1}\dot{\rho}_{*, {\rm LBG}}(z^\prime)\frac{dt}{dz^\prime}dz^\prime \;,
\label{eq:rhostar}
\end{equation}

\noindent where 

\begin{equation}
\frac{dt}{dz^\prime}=\frac{1}{(1+z^\prime)H(z^\prime)} \;,
\label{eq:dtdz}
\end{equation}
$\dot{\rho_{*}}$ is corrected for extinction, and ${z^\prime_2}$ and ${z^\prime_1}$ represent redshift intervals at which $\dot{\rho_{*}}$ is measured.
While $\rho_{\rm metals}(z)_{\rm LBG}$ is a differential quantity for each redshift interval, the steep rise in $\dot{\rho_{*}}$ with decreasing redshift results in the cumulative 
$\rho_{\rm metals}(z)_{\rm LBG}$ from $z=10$ being equivalent within the errors. 

These are just lower limits to $\rho_{\rm metals}(z)_{\rm LBG}$ as we only integrate the luminosity function out to $0.04-0.05$L$^*$, and we must also consider the metal contribution of the faint end. 
Recent measurements of the faint end of the luminosity function at $z\sim2$ \citep{Alavi:2014} suggest a  significant contribution to $\dot{\rho}_{*, {\rm LBG}}$ from faint low mass galaxies, suggesting we may be missing as much as half the metals, depending on the effects of dust and redshift. This suggests that LBGs produce an even larger fraction of the total metals in the Universe.
Our calculations assume that the ratio of $\rho_{\rm metals}(z)_{\rm U}$ to $\rho_{\rm metals}(z)_{\rm LBG}$ does not evolve significantly with redshift. This means we are assuming that LBGs still dominate the metal production at high redshift, and that the metal production rate for a given $\dot{\rho_*}$ does not significantly depend on redshift. 

\begin{figure}[]
\center{
\includegraphics[scale=0.5, viewport=15 5 495 360,clip]{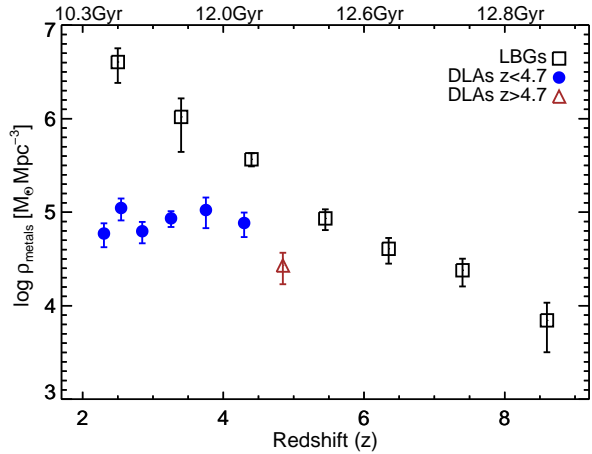}
}
\caption{The mass density of metals as a function of redshift for DLAs and LBGs per redshift interval.
The black squares are $\rho_{\rm metals}(z)$ produced by LBGs, the blue filled circles are $\rho_{\rm metals}(z)$ for DLAs, and the brown triangle
is $\rho_{\rm metals}(z)$ for DLAs at $z>4.7$. This shows that DLAs are an important contributor to the total metal budget at $z>4$,
and that $\rho_{\rm metals}(z)$ may begin to decrease at $z>4.7$.
}
\label{fig:rho}
\end{figure}

We compare $\rho_{\rm metals}(z)_{\rm LBG}$ to $\rho_{\rm metals}(z)_{\rm DLAs}$ in Figure \ref{fig:rho}.
The increasing $\rho_{\rm HI}$  and decreasing {\Z} of DLAs results in a relatively flat $\rho_{\rm metals}(z)_{\rm DLAs}$,
while the  $\rho_{\rm metals}(z)_{\rm LBG}$ decreases rapidly at higher redshifts. 
At $z\sim2.3$, DLAs comprise only $\sim1\%$ of the metals produced by LBGs, but this fraction grows to $\sim20$\% by $z\sim4.3$.
These ratios could be up to a factor of two overestimated due to the faint end of the luminosity function described above. 
Regardless, this suggests that the diffuse HI gas is significantly enriched by the star-forming regions of LBGs soon after reionization, or are forming a significant fraction of metals in-situ.

One intriguing consequence of the decrease in metallicity of DLAs at $z>4.7$ is its effect on $\rho_{\rm metals}(z)_{\rm DLA}$. 
The brown triangle in Figure \ref{fig:rho} represents $\rho_{\rm metals}(z)_{\rm DLA}$ at $z\sim4.85$, which appears to decrease for the first time since $z\sim2$. 
We are therefore likely measuring the beginning of a decrease in the density of metals in neutral gas as we approach the epoch of reionization. 

\section{Discussion}
\label{discus}

The principal result of this study is the first measurement of the cosmic metallicity {\Z} of neutral gas at $z\sim5$ of $\langle {\rm Z}\rangle=-2.03^{+0.09}_{-0.11}$. This measurement provides evidence of rapid metallicity enrichment  of DLAs just after the reionization epoch, with a $6\sigma$ deviation of {\Z} from the extrapolated linear trend at $z>4.7$. We also find that the contribution of DLAs to the total metal budget significantly increases with redshift, emphasizing the importance of the sudden metallicity evolution.

A possible explanation for the observed decrease in metallicity at the highest redshifts is that the covering fraction of neutral gas outside disks increases as a function of redshift. The combined effects of an increase in the density of the universe and a decrease in the background radiation field \citep{Haardt:2012} enable hydrogen to self-shield at densities below the virial density at high redshifts \citep{McQuinn:2011}. Therefore, neutral hydrogen may be found more frequently in the halos of galaxies \citep[e.g. in cold flows; ][]{Keres:2005,Dekel:2006}, or in the denser regions of the intergalactic medium \citep{McQuinn:2011, Fumagalli:2013a}, creating a second population of DLAs \citep[e.g.][]{Berry:2013}. If cold flows or the intergalactic medium have a lower level of enrichment compared to the interstellar medium  \citep[e.g.][]{Fumagalli:2011b},
then the average metallicity in DLAs would rapidly decline with redshift as this second DLA population becomes more and more predominant. Given the age of the universe at $z\gtrsim5$, this interpretation implies that metals need to be transported into the halos on short time scales, and have to either arrive in a neutral state, or recombine quickly. 

Besides this hypothesis, other explanations can also produce the observed change in the rate of metal enrichment in DLAs, such as a change in the physical processes that enrich neutral gas within disks. Future simulations and models will need to test more quantitatively the feasibility of these possibilities.

\acknowledgements

We thank Rob Simcoe for sharing a FIRE spectrum of J0824+1302 in advance of publication; 
part of these data were obtained as part of observations supported by NSF award AST-1109915.
Support for this work was provided by NSF grant AST-1109447.
MF is supported by a Hubble Fellowship grant HF-51305.01-A.
This paper includes data gathered at the W.M. Keck Observatory and Las Campanas Observatory.

{\it Facility:} 
\facility{Keck:II (ESI)}


\end{document}